\documentclass[prl,12pt,showpacs,amsmath,onecolumn,floatfix]{revtex4}
\usepackage{amsmath}
\usepackage{graphicx}

\newcommand{\ket}[1]{|#1\rangle}
\newcommand{\bra}[1]{\langle#1|}
\newcommand{\avg}[1]{\langle#1\rangle}
\newcommand{\Avg}[1]{\left\langle#1\right\rangle}

\newcommand{\bs}[1]{\boldsymbol{#1}}
\newcommand{\sphere}[1]{\Omega_{#1},\theta_{#1},\phi_{#1},\sigma_{#1}}
\begin{document}
\title{Ultimate Energy Densities for Electromagnetic Pulses}
\author{Mankei Tsang}
\email{mankei@mit.edu}
\affiliation{Research Laboratory of Electronics,
Massachusetts Institute of Technology,
Cambridge, Massachusetts 02139, USA}

\date{\today}

\begin{abstract}
  The ultimate electric and magnetic energy densities that can be
  attained by bandlimited electromagnetic pulses in free space are
  calculated using an \textit{ab initio} quantized treatment, and the
  quantum states of electromagnetic fields that achieve the ultimate
  energy densities are derived. The ultimate energy densities also
  provide an experimentally accessible metric for the degree of
  localization of polychromatic photons.
\end{abstract}

\maketitle
Ultrafast lasers have become an indispensible tool in a wide spectrum
of science, including nonlinear optics \cite{brabec,mourou}, metrology
\cite{cundiff}, laser fusion \cite{tabak}, biological imaging
\cite{zipfel}, biological surgery \cite{juhasz}, and chemistry
\cite{zewail}.  A key to the success of such lasers is the extremely
high peak energy density that they can achieve, as the moderate energy
of each laser pulse can be confined within femtoseconds or even
attoseconds and focused to a micron-sized area. The high energy
density strongly enhances light-matter interactions, and is especially
crucial to the study of relativistic nonlinear optics
\cite{mourou}. Given the importance of an ultrahigh optical energy
density in a broad range of applications, the limit to which one can
focus a broadband optical pulse in three spatiotemporal dimensions and
maximize the energy density is a fundamental problem.

Localization of electromagnetic pulses has been investigated both in
the classical regime \cite{ziolkowski} and the quantum single-photon
regime \cite{amrein,chan}. While these seminal studies have
contributed to our fundamental understanding of electromagnetic energy
localization, all of their predictions have not yet been
experimentally verified because of the difficulty in implementing
their proposed electromagnetic pulse solutions.  Most require a
spectrum that contains arbitrarily high frequency components
\cite{ziolkowski,amrein} and can be exceedingly difficult to realize
due to the finite laser gain bandwidth or finite transparency range of
optical components. The use of a spontaneously emitting atom proposed
in Ref.~\cite{chan} couples the properties of the emitted photon to
those of the atom and does not seem to be generalizable to other
situations. Moreover, the quantum studies \cite{amrein,chan} are
mainly concerned with the decay of the energy density far away from
the center of localization at an instant of time for one photon, and
thus are not immediately relevant to the more practical problem of
maximizing the energy density at the center of localization for a
large number of bandlimited photons.

In this Letter, the ultimate electric and magnetic energy densities
that can be attained by bandlimited electromagnetic pulses in free
space are calculated using an \textit{ab initio} quantized treatment,
and the quantum states that achieve the ultimate densities are
derived. By taking into account all degrees of freedom of
electromagnetic fields and explicitly limiting the bandwidth of the
pulses, our result overcomes all the shortcomings of the
aforementioned studies and is more applicable to experimental
situations. Measuring the energy densities at a fixed point is also
considerably easier experimentally than measuring the decay of the
energy density at an instant of time, so the maximum achievable energy
densities can be used as an alternative and more accessible metric for
the degree of localization of polychromatic photons. Most importantly,
the ultimate energy densities impose a fundamental limit to which a
bandlimited optical pulse can be focused spatially and temporally, so
the presented result should prove useful for designing ultrafast optics
experiments.

The procedure of calculating the maximum energy densities is similar
to the one used to calculate the multiphoton absorption rate limit for
monochromatic light in Ref.~\cite{tsang}, except that here we
generalize the procedure to polychromatic light, such that all degrees
of freedom are taken into account and the treatment can be considered
\textit{ab initio}. We also calculate the maximum magnetic energy
density and the corresponding quantum state, as the magnetic field can
also play a significant role in relativistic nonlinear optics
\cite{mourou}.

We first derive the ultimate electric energy density, since it is more
important for most applications.  Consider the quantized electric
field operator in free space \cite{mandel}:
\begin{align}
\hat{\bs{E}}(\bs{r},t)&=
\frac{i}{(2\pi)^{3/2}}\sum_\sigma\int d^3k
\left(\frac{\hbar\omega}{2\epsilon_0}\right)^{1/2}
\bs\varepsilon(\bs{k},\sigma)
\hat{a}(\bs{k},\sigma)e^{i\bs{k}\cdot\bs{r}-i\omega t} +\textrm{H.c.},
\end{align}
where $\sigma$ denotes the two transverse polarizations,
$\bs\varepsilon(\bs{k},\sigma)$ is the unit electric-field
polarization vector, $\omega = ck = c(k_x^2+k_y^2+k_z^2)^{1/2}$ is the
frequency, $\hat{a}(\bs{k},\sigma)$ is the annihilation operator
satisfying the commutation relation $[\hat{a}(\bs{k},\sigma),
\hat{a}^\dagger(\bs{k}',\sigma')] =
\delta^3(\bs{k}-\bs{k}')\delta_{\sigma\sigma'}$, and H.c. is the
Hermitian conjugate.  To impose a limit on the bandwidth, it is
necessary to describe the optical modes in terms of the frequency
variable. This can be done by changing the momentum-space coordinates
$(k_x,k_y,k_z)$ to normalized spherical coordinates
$(\Omega,\theta,\phi)$:
\begin{align}
\omega &= \omega_0 \Omega,
&\quad
k_x &= k_0\Omega\sin\theta\cos\phi,
\nonumber\\
k_y &= k_0\Omega\sin\theta\sin\phi,
&\quad
k_z &= k_0\Omega\cos\theta,
\nonumber\\
dk_xdk_ydk_z &= d\Omega d\theta d\phi \left(k_0^3\Omega^2\sin\theta\right),
&\quad
\hat{a}(\bs{k},\sigma) &=
\hat{a}(\Omega,\theta,\phi,\sigma)
\left(k_0^3\Omega^2\sin\theta\right)^{-1/2}.
\end{align}
where $\omega_0$ is a normalization frequency, $k_0 \equiv \omega_0/c
\equiv 2\pi/\lambda_0$, and the annihilation operator has been
renormalized so that its commutator is
$[\hat{a}(\Omega,\theta,\phi,\sigma),
\hat{a}^\dagger(\Omega',\theta',\phi',\sigma')]
=\delta(\Omega-\Omega')\delta(\theta-\theta')\delta(\phi-\phi')
\delta_{\sigma\sigma'}$. The positive-frequency electric field becomes
\begin{align}
\hat{\bs{E}}^{(+)}(\bs{r},t) &=
i\left(\frac{\hbar\omega_0}{2\epsilon_0\lambda_0^3}\right)^{1/2}
\int_0^\infty d\Omega\int_0^\pi d\theta\int_0^{2\pi}d\phi
\left(\Omega^3\sin\theta\right)^{1/2}
\bs\varepsilon(\theta,\phi,\sigma)
\hat{a}(\Omega,\theta,\phi,\sigma)
e^{i\bs{k}\cdot\bs{r}-i\omega t}.
\end{align}

In the continuous Fock space representation \cite{mandel}, the
$N$-photon momentum eigenstate is given by
\begin{align}
\ket{\sphere{1},\dots,\sphere{N}} 
&\equiv
\frac{1}{\sqrt{N!}}\prod_{n=1}^N
\hat{a}^\dagger(\sphere{n})\ket{0},
\end{align}
and the identity operator is
\begin{align}
\hat{1} &=
\sum_{N=0}^\infty\ket{N}\bra{N},
\\
\ket{N}\bra{N}&=\sum_{\sigma_1,\dots,\sigma_N}
\int d\Omega_1d\theta_1d\phi_1\dots d\Omega_Nd\theta_Nd\phi_N
\nonumber\\&\quad\times
\ket{\sphere{1},\dots,\sphere{N}} 
\bra{\sphere{1},\dots,\sphere{N}}.
\end{align}
An arbitrary quantum state of electromagnetic
fields can thus be expressed as
\begin{align}
\ket{\Psi} &= \sum_{N=0}^\infty C_N\ket{N},
\quad C_N \equiv \Avg{N|\Psi},
\end{align}
and a Fock state as
\begin{align}
\ket{N} &=\sum_{\sigma_1,\dots,\sigma_N}
\int d\Omega_1d\theta_1d\phi_1\dots d\Omega_Nd\theta_Nd\phi_N
\Phi_N(\sphere{1},\dots,\sphere{N})
\nonumber\\&\quad\times
\ket{\sphere{1},\dots,\sphere{N}},
\end{align}
where
\begin{align}
\Phi_N(\sphere{1},\dots,\sphere{N})
&\equiv\bra{\sphere{1},\dots,\sphere{N}}N\rangle
\end{align}
is the $N$-photon momentum-space probability amplitude, which
must satisfy the normalization condition:
\begin{align}
&\sum_{\sigma_1,\dots,\sigma_N}
\int d\Omega_1d\theta_1d\phi_1\dots d\Omega_Nd\theta_Nd\phi_N
\left|\Phi_N(\sphere{1},\dots,\sphere{N})\right|^2 = 1,
\label{norm}
\end{align}
and the symmetrization condition:
\begin{align}
&\quad\Phi_N(\dots,\sphere{n},\dots,\sphere{m},\dots)
\nonumber\\&=
\Phi_N(\dots,\sphere{m},\dots,\sphere{n},\dots)
\textrm{ for any }n \textrm{ and }m.
\label{symmetric}
\end{align}
To impose a limited bandwidth ($\alpha \le \Omega \le \beta$) on the
electromagnetic fields, we require the probability of photons existing
outside the bandwidth to vanish:
\begin{align}
\Phi_N(\sphere{1},\dots,\sphere{N}) = 0
\textrm{ for any }\Omega_n < \alpha \textrm{ or } \Omega_n > \beta.
\end{align}

With the theoretical framework put forth, we now proceed to calculate
a bound on the electric energy density (minus the zero-point energy
density) given by
\begin{align}
U_e &\equiv \Avg{:\frac{\epsilon_0}{2}\hat{\bs{E}}
\cdot\hat{\bs{E}}:}=
\Avg{\epsilon_0\hat{\bs{E}}^{(-)}\cdot\hat{\bs{E}}^{(+)}}.
\end{align}
A bound on the electric energy density is equivalent to
a bound on the energy density for one component of the electric field:
\begin{align}
U_{e}' &\equiv 
\Avg{:\frac{\epsilon_0}{2}\left(\bs{p}\cdot\hat{\bs{E}}\right)^2:}
=\epsilon_0 \Avg{\left(\bs{p}\cdot \hat{\bs{E}}^{(-)}\right)
\left(\bs{p}\cdot \hat{\bs{E}}^{(+)}\right)},
\end{align}
where $\bs{p}$ is an arbitrary real unit vector, because $U_e$ and
$U_e'$ are equivalent if we choose $\bs{p}$ to be parallel to the
electric field. In terms of the momentum-space representation,
\begin{align}
U_{e}' &=\frac{\hbar\omega_0}{2\lambda_0^3}
\sum_{N=0}^\infty |C_N|^2N
\sum_{\sigma_2,\dots,\sigma_N}\int d\Omega_2 d\theta_2d\phi_2\dots
d\Omega_Nd\theta_Nd\phi_N
\nonumber\\&\quad\times
\Bigg|\sum_\sigma\int_{\alpha}^\beta d\Omega
\int_0^\pi d\theta\int_0^{2\pi}d\phi
\left[i\left(\Omega^3\sin\theta\right)^{1/2}
\bs{p}\cdot\bs\varepsilon(\theta,\phi,\sigma)
e^{i\bs{k}\cdot\bs{r}-i\omega t}\right]
\nonumber\\&\quad\times
\Phi_N(\Omega,\theta,\phi,\sigma,\sphere{2},
\dots,\sphere{N})\Bigg|^2,
\label{inner}
\end{align}
where the symmetric property of $\Phi_N$ given by
Eq.~(\ref{symmetric}) is used.  By virtue of the Schwarz's inequality
and the normalization condition given by Eq.~(\ref{norm}),
\begin{align}
U_{e}' &\le \frac{\hbar\omega_0}{2\lambda_0^3}
\sum_{N=0}^\infty |C_N|^2N
\sum_{\sigma_1,\dots,\sigma_N}\int d\Omega_1 d\theta_1d\phi_1\dots
d\Omega_Nd\theta_Nd\phi_N
\nonumber\\&\quad\times
 |\Phi_N(\sphere{1},\dots,\sphere{N})|^2
\nonumber\\&\quad\times
\sum_\sigma \int_{\alpha}^\beta d\Omega
\int_0^\pi d\theta\int_0^{2\pi}d\phi
\left(\Omega^3\sin\theta\right)
\left|\bs{p} \cdot\bs\varepsilon(\theta,\phi,\sigma)\right|^2
\nonumber\\
&=\frac{\pi}{3}\frac{\avg{N}\hbar\omega_0}{\lambda_0^3}
\left(\beta^4-\alpha^4\right),
\end{align}
where $\avg{N}\equiv \sum_N|C_N|^2N$ is the average photon number. As
expected, the bound on $U_e'$ does not depend on $\bs{p}$, and is
therefore also applicable to the total electric energy density.  Defining
the actual lower and upper frequencies as $\omega_1 = \alpha\omega_0$
and $\omega_2 = \beta\omega_0$, respectively, and the corresponding
wavelengths as $\lambda_{1,2} = 2\pi c/\omega_{1,2}$, we obtain the
central result of this Letter:
\begin{align}
\Avg{:\frac{\epsilon_0}{2}\hat{\bs{E}}\cdot\hat{\bs{E}}:}
\le \frac{\pi}{3}\Avg{N}
\left(\frac{\hbar\omega_2}{\lambda_2^3}
-\frac{\hbar\omega_1}{\lambda_1^3}\right).
\label{bound_electric}
\end{align}
This simple expression agrees with the intuition that the ultimate
energy density is limited by the maximum energy of photons
($\Avg{N}\hbar\omega_2$) divided by the smallest volume that the
photons can occupy ($\lambda_2^3$). 

In the limit of a small bandwidth compared to the center frequency, we
can let $\Delta\omega \equiv \omega_2-\omega_1$, $\omega_0 =
(\omega_1+\omega_2)/2$, $\Delta\omega \ll \omega_0$, and obtain
\begin{align}
\Avg{:\frac{\epsilon_0c}{2}\hat{\bs{E}}\cdot\hat{\bs{E}}:}
\lesssim \frac{2}{3}\Avg{N}
\frac{\hbar\omega_0\Delta\omega}{\lambda_0^2},
\end{align}
which is a bound on the peak intensity in the slowly-varying envelope
regime, and again agrees with the intuition that the highest intensity
is achieved when the mean energy of the photons is focused to their
minimum pulse width ($2\pi/\Delta\omega$) and beam size
($\lambda_0^2$). This approximate bound also agrees with that derived
in Ref.~\cite{tsang}, where the monochromatic approximation is made at
the beginning.  Beyond the slowly-varying envelope regime, the exact
bound given by Eq.~(\ref{bound_electric}) depends on the upper
frequency to the fourth power, underlying the importance of
high-frequency components in maximizing the energy density, as they
have a higher energy as well as a smaller localization volume.

The use of the Schwarz's inequality is reminiscent of the matched
filter concept in communication theory \cite{couch}.  In
Eq.~(\ref{inner}), the $N$-photon amplitude can be regarded as the
input signal, and the expression in square brackets can be regarded as
a filter transfer function in the measurement of the electric
field. An $N$-photon amplitude that achieves the Schwarz upper bound
is one that is linearly dependent on the square-bracketed
expression, or in other words, when the input signal matches the
filter. Assuming a factorizable $\Phi_N$, the following $N$-photon
amplitude that achieves the ultimate electric energy density can then
be obtained:
\begin{align}
\Phi_N &= \prod_{n=1}^N f_e(\Omega_n,\theta_n,\phi_n,\sigma_n),
\nonumber\\
f_e &=\bigg\{\begin{array}{l}
-iC^{-1/2}(\Omega^3\sin\theta)^{1/2}
\bs{p}\cdot\bs\varepsilon^*(\theta,\phi,\sigma)
e^{-i\bs{k}\cdot\bs{r}_0+i\omega t_0}
\textrm{ for }\alpha \le \Omega_n \le \beta,
\\
0 \textrm{ otherwise,}
\end{array}
\nonumber\\
C&= \frac{2\pi}{3}\left(\beta^4-\alpha^4\right).
\label{factorizable}
\end{align}
This state produces the maximum electric energy density at
$(\bs{r}_0,t_0)$, with the electric field at $(\bs{r}_0,t_0)$
polarized along $\bs{p}$. 

To apply the above result to the classical regime, let
\begin{align}
C_N &= e^{-\avg{N}/2}\frac{\avg{N}^{N/2}}{\sqrt{N!}}.
\end{align}
Together with a factorizable $\Phi_N$ in Eq.~(\ref{factorizable}),
the quantum state becomes a coherent state in the continuous
mode representation \cite{mandel}, and the mean electric field is
then given by
\begin{align}
\bs{E}(\bs{r},t) &=
i\left(\frac{\Avg{N}\hbar\omega_0}{2\epsilon_0\lambda_0^3}\right)^{1/2}
\int_\alpha^\beta d\Omega\int_0^\pi d\theta\int_0^{2\pi}d\phi
\nonumber\\&\quad\times
\left(\Omega^3\sin\theta\right)^{1/2}
\bs\varepsilon(\theta,\phi,\sigma)
f_e(\Omega,\theta,\phi,\sigma)e^{i\bs{k}\cdot\bs{r}-i\omega t} +
\textrm{H.c.},
\end{align}
which, incidentally, must be an exact solution of the Maxwell
equations.  The Fourier transform of $\bs{E}(\bs{r},t)$ is
proportional to $\omega^3$, and the classical power spectrum is then
proportional to $\omega^6$ within the allowed frequency band.

Consider now the magnetic field operator:
\begin{align}
\hat{\bs{B}}(\bs{r},t)
&=\frac{i}{(2\pi)^{3/2}}\sum_\sigma\int d^3k
\left(\frac{\mu_0\hbar\omega}{2}\right)^{1/2}
\bs\kappa\times \bs\varepsilon(\bs{k},\sigma)
\hat{a}(\bs{k},\sigma)e^{i\bs{k}\cdot\bs{r}-i\omega t} +\textrm{H.c.},
\end{align}
where $\bs\kappa \equiv \bs{k}/k$. While the total magnetic energy
must be the same as the total electric energy for photons in free
space, it is not difficult to show that the magnetic energy density at
$(\bs{r}_0,t_0)$ is zero where the electric energy density is maximum
for the state given by Eqs.~(\ref{factorizable}). To maximize the
magnetic energy density instead, we can simply apply the same
procedure as above to the magnetic energy density, which turns out to
obey the same bound as the electric one:
\begin{align}
\Avg{:\frac{1}{2\mu_0}\hat{\bs{B}}\cdot\hat{\bs{B}}:}
\le \frac{\pi}{3}\Avg{N}
\left(\frac{\hbar\omega_2}{\lambda_2^3}
-\frac{\hbar\omega_1}{\lambda_1^3}\right).
\end{align}
The quantum state with the ultimate magnetic energy density is also
similar to the electric case,
\begin{align}
\Phi_N &= \prod_{n=1}^N f_b(\Omega_n,\theta_n,\phi_n,\sigma_n),
\nonumber\\
f_b &=\bigg\{\begin{array}{l}
-iC^{-1/2}(\Omega^3\sin\theta)^{1/2}
\bs{m}\cdot[\bs\kappa\times\bs\varepsilon^*(\theta,\phi,\sigma)]
e^{-i\bs{k}\cdot\bs{r}_0+i\omega t_0}
\textrm{ for }\alpha \le \Omega_n \le \beta,
\\
0 \textrm{ otherwise,}
\end{array}
\end{align}
where $\bs{m}$ is the unit vector of the magnetic field at
$(\bs{r}_0,t_0)$. 

The ultimate electric and magnetic energy densities may be challenging
to achieve experimentally, as they require a power spectrum
proportional to $\omega^6$ within the allowed band, spatial focusing
in all directions with a specific angular spectrum, and polarization
control. That said, the results set forth impose fundamental limits to
which the energy densities can reach regardless of technological
advances in the control of electromagnetic fields, and therefore
should prove useful for designing ultrafast optics experiments.

The author would like to acknowledge financial support
from the William M.\ Keck Foundation Center for Extreme
Quantum Information Theory.


\begin{thebibliography}{}
\bibitem{brabec}T.\ Brabec and F.\ Krausz,
\rmp \textbf{72}, 545 (2000).

\bibitem{mourou} G.\ A.\ Mourou, T.\ Tajima, and S.\ V.\ Bulanov,
\rmp \textbf{78}, 309 (2006).

\bibitem{cundiff} S.\ T.\ Cundiff and J.\ Ye,
\rmp \textbf{75}, 325 (2003).

\bibitem{tabak}M.\ Tabak \textit{et al.},
Phys.\ Plasmas \textbf{1}, 1626 (1994).


\bibitem{zipfel}W.\ Denk, J.\ H.\ Strickler, and W.\ W.\ Webb,
Science \textbf{248}, 73 (1990);
W.\ R.\ Zipfel, R.\ M.\ Williams, W.\ W.\ Webb,
Nat.\ Biotechnol.\ \textbf{11}, 1369 (2003).

\bibitem{juhasz} T.\ Juhasz, F.\ H.\ Loesel, R.\ M.\ Kurtz, C.\ Horvath,
J.\ F.\ Bille, and G.\ A.\ Mourou,
IEEE J.\ Sel.\ Topics Quantum Electron.\ \textbf{5}, 902 (1999).

\bibitem{zewail}A.\ H.\ Zewail,
J.\ Phys.\ Chem.\ A \textbf{104}, 5660 (2000).

\bibitem{ziolkowski}R.\ W.\ Ziolkowski,
\pra \textbf{39}, 2005 (1989);
R.\ W.\ Hellwarth and P.\ Nouchi,
\pre \textbf{54}, (1996).

\bibitem{amrein}W. O. Amrein,
Helv.\ Phys.\ Acta \textbf{8}, 2684 (1969);
C.\ Adlard, E.\ R.\ Pike, and S.\ Sarkar,
\prl \textbf{79}, 1585 (1997);
I.\ Bialynicki-Birula,
\prl \textbf{80}, 5247 (1998).

\bibitem{chan} K.\ W.\ Chan, C.\ K.\ Law, and J.\ H.\ Eberly,
\prl \textbf{88}, 100402 (2002).

\bibitem{tsang}M.\ Tsang, e-print arXiv:0802.0516v1.

\bibitem{mandel}L.\ Mandel and E.\ Wolf,
\textit{Optical Coherence and Quantum Optics}
(Cambridge University Press, Cambridge, UK, 1995).

\bibitem{couch}L.\ W.\ Couch,
\textit{Digital and Analog Communication Systems}
(Prentice Hall, New Jersey, 2001).



\end{thebibliography}
\end{document}